\documentclass[12pt]{reportj}
\usepackage{deluxetablej}
\usepackage{hyperref} 
\makeatletter
\def\@to{to}
\makeatother
\usepackage{times}
\usepackage{graphicx}
\usepackage{xcolor}
\usepackage{tocloft}
\usepackage{fancyheadings}
\emergencystretch=\maxdimen
\def\code#1{\texttt{#1}}
\renewcommand\thesection{\arabic{section}}
\renewcommand\thesubsection{\thesection.\arabic{subsection}}

\def\ssection#1{\setcounter{subsection}{0} \refstepcounter{section} \section*{\hbox to \hsize{\large\bf \arabic{section}. #1\hfill }}\label{sec} \addcontentsline{toc}{section}{\arabic{section}. #1}}
\def\ssubsection#1{\setcounter{subsubsection}{0} \refstepcounter{subsection}\subsection*{\hbox to \hsize{\normalsize\bfseries\itshape \arabic{section}.\arabic{subsection} #1\hfill}}\label{subsec} \addcontentsline{toc}{subsection}{\arabic{section}.\arabic{subsection} #1}}
\def\ssubsubsection#1{\refstepcounter{subsubsection}\subsection*{\hbox to \hsize{\normalsize\it \arabic{section}.\arabic{subsection}.\arabic{subsubsection} #1\hfill}}\label{subsubsec} \addcontentsline{toc}{subsubsection}{\arabic{section}.\arabic{subsection}.\arabic{subsubsection} #1}}

\def\ssectionstar#1{\section*{\hbox to \hsize{\large\bf #1\hfill}} \addcontentsline{toc}{section}{#1}}
\def\ssubsectionstar#1{\subsection*{\hbox to \hsize{\normalsize\bfseries\itshape #1\hfill}} \addcontentsline{toc}{subsection}{#1}}
\def\ssubsubsectionstar#1{\subsection*{\hbox to \hsize{\normalsize\it  #1\hfill}} \addcontentsline{toc}{subsection}{#1}}

\renewcommand{\cftaftertoctitle}{%
\mbox{}\hfill{\normalfont Page}}
\begin{document}

~\\

\vspace{-2.4cm}
\noindent\includegraphics*[width=0.295\linewidth]{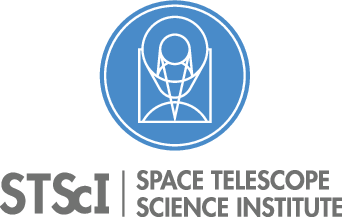}

\vspace{-0.4cm}

\begin{flushright}
    {\bf Instrument Science Report STIS 2024-03}
    
    \vspace{1.1cm}
    
    {\bf\Huge Rederivation of STIS Secondary Echelle Mode Traces}
    
    \rule{0.25\linewidth}{0.5pt}
    
    \vspace{0.5cm}
    
    Matthew R. Siebert $^1$, TalaWanda Monroe $^1$, and Svea Hernandez $^1$
    \linebreak
    \newline
    \footnotesize{$^1$ Space Telescope Science Institute, Baltimore, MD\\}
    
    \vspace{0.5cm}
    
     10 June 2024
\end{flushright}

\vspace{0.1cm}

\noindent\rule{\linewidth}{1.0pt}
\noindent{\bf A{\footnotesize BSTRACT}}

{\it \noindent The STIS echelle gratings can be used with a variety of different central wavelength settings. ``Secondary" wavelength settings, designed to cover select absorption or emission lines, have not been calibrated as precisely as their primary mode counterparts. In particular, secondary echelle mode traces (and subsequent extraction regions) have been previously defined using straight line fits to each spectral order. In this work, we define a new general method for defining echelle traces that utilizes Gaussian process regression and accounts for the detailed curvature of each order across the detector. Across a variety of echelle grating and central wavelength settings, we find that this method can improve flux throughput by $\sim\!\!\!\!\!4\%$ especially near wavelengths located close to the edge of the detector. We have used this method to provide new traces and update reference files for 9 different echelle modes for both pre- and post-Servicing Mission 4 (SM4; in 2009) observations.}

\vspace{-0.1cm}
\noindent\rule{\linewidth}{1.0pt}

\renewcommand{\cftaftertoctitle}{\thispagestyle{fancy}}
\tableofcontents


\vspace{-0.3cm}
\ssection{Introduction}\label{sec:Introduction}

\noindent The Space Telescope Imaging Spectrograph has four echelle gratings, E140H, E140M, E230H, and E230M, that provide the unique capability of observing astronomical sources at UV wavelengths ($\sim 1100 - 3100$\AA) with medium to high resolution ($R\sim 30{,}000 - 110{,}000$). Each of these gratings have central wavelength settings that have been designated as ``primary" observing modes, designed to span the spectral range of the grating. In addition to these primary modes, three of these gratings (E140H, E230H, and E230M) offer ``secondary" observing modes using central wavelengths that are better centered on absorption or emission lines of interest. The E140H, E230H, and E230M gratings offer 8, 20, and 4 secondary central wavelengths, respectively. 

According to the \href{https://hst-docs.stsci.edu/stisihb}{STIS Instrument Handbook} (Medallon, Welty et al., 2023), the absolute and relative flux accuracies for all echelle settings are of the order 8\% and 5\%, respectively. Recent improvements to primary standard star white dwarf atmospheric models (\href{https://ui.adsabs.harvard.edu/abs/2020AJ....160...21B/abstract}{Bohlin et al.\ 2020)} have shown roughly 1-3\% continuum differences from past work. As a result of these updates, the STIS team has an ongoing flux-recalibration effort aimed at deriving new sensitivity curves and updating other relevant reference files for a variety of STIS observing modes using these new models (CALSPECv11). These updates have prioritized the most used STIS observing modes and so far, for the echelle gratings, the team has produced new sensitivity functions and blaze shift coefficients for E140M/1425, E230M/1978, E230M/2707, E230M/2415 (secondary), E230H/2263, and E230H/2713 (see STIS STANs for April 2022\footnote{\url{https://www.stsci.edu/contents/news/stis-stans/april-2022-stan\#article1}}, January 2023\footnote{\url{https://www.stsci.edu/contents/news/stis-stans/january-2023-stan\#article1}}, and July 2023\footnote{\url{https://www.stsci.edu/contents/news/stis-stans/july-2023-stan\#fluxrecal}}).

Several echelle secondary observing modes are used as frequently as the primary modes of the same grating. For example, E140H/1234 (primary), E140H/1271 (secondary), and E140H/1307 (secondary) are each used for roughly 10 observations per year. Given the consistent usage of some of these secondary modes, the ongoing flux recalibration work included updates to their throughputs to maintain the nominal flux accuracies. One particular component to improving the flux calibration of these secondary modes involves the rederivation of their traces. The trace of each echelle order maps the location of the spectrum on the detector as a function of wavelength and is used in the spectral extraction step. For primary echelle modes, pre-defined traces follow the slightly curved paths of the dispersed light across the detector. However, echelle secondary traces have historically been defined using straight-line fits to each order. The work in this ISR is primarily focused on understanding the impact and potential improvement from updating echelle secondary mode traces to more optimally characterize the shape of their individual orders. 

\ssubsection{Reference File Updates}
Echelle trace information is stored in the SPTRCTAB reference files. The shape of each trace is stored as an array (A2DISPL) consisting of offsets in pixels in the Y direction relative to A2CENTER (the Y position of the trace at the central column of the detector) and SHIFTA2 (the Y-offset from the nominal placement of the image due to differences in Mode Selection Mechanism settings).

\begin{figure}[htb!]
  \centering
  \includegraphics[width=5.8in]{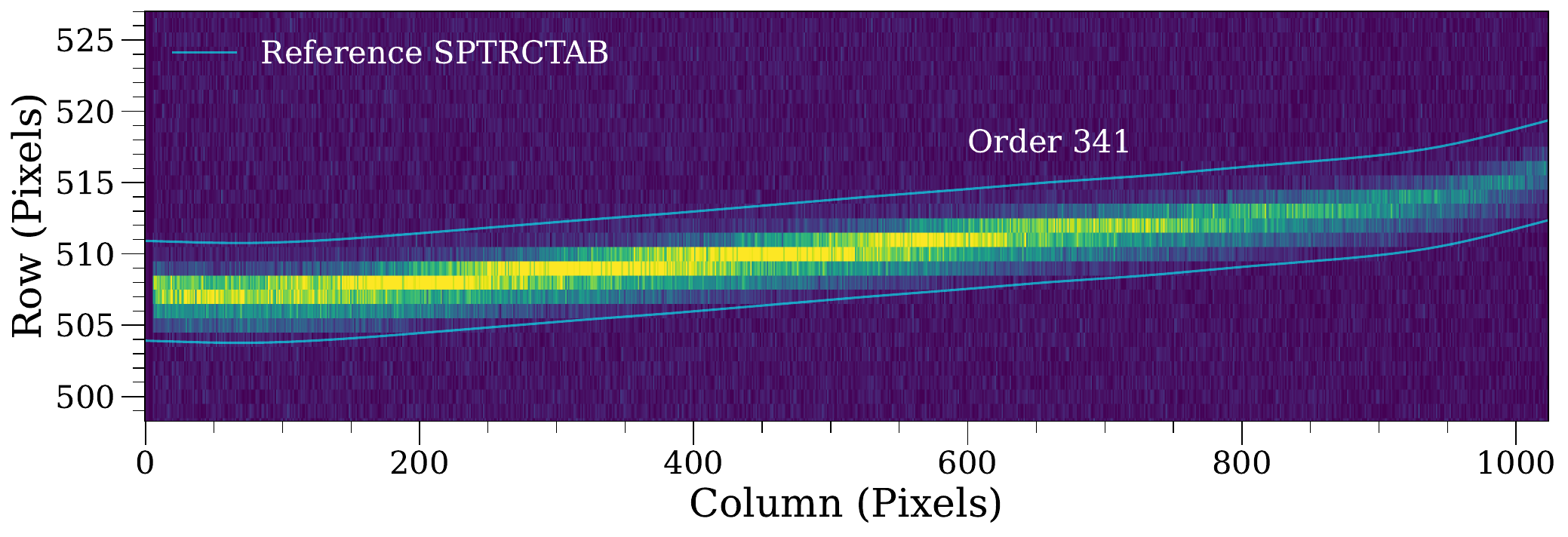}
  \includegraphics[width=5.8in]{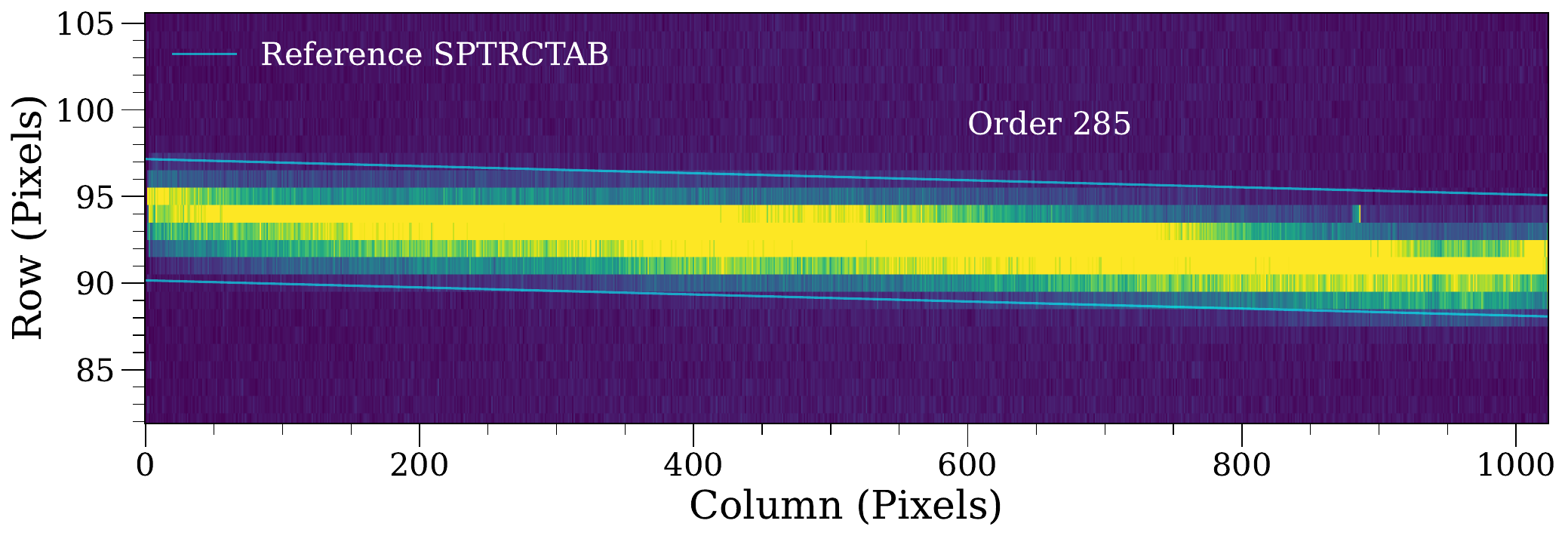}
    \caption{Example extraction regions of select orders taken from a primary echelle mode (top, E230H/2263) and a secondary echelle mode (bottom, E140H/1562). In the former example, the curvature of the order is reflected in the shape of the trace, however, in the latter example, the straight line trace is clearly insufficient in centering the cross-dispersion profile in the extraction region. This leads to missing flux, especially near the edges of the detector.}
    \label{fig:trace}
\end{figure}

In \autoref{fig:trace} we present echelle observations of the standard star G191-B2B. The top panel shows a 2D-cutout image centered on order 341 of the primary mode E230H/2263, and the bottom panel shows a 2D-cutout image centered on order 285 of the secondary mode E140H/1562. The blue lines represent the pre-defined spectral traces, i.e., extraction regions used in the pipeline reduction. It is clear that the primary mode trace better represents the curvature of the order and how the trace behaves near the edges of the detector. The secondary mode trace is not well-centered and is likely missing flux at the edges. These differences are representative of the trends seen in echelle secondary traces across all gratings.

While many reduction pipelines typically derive traces specifically for individual observations, CALSTIS uses pre-defined spectral trace shapes to perform extraction of all spectroscopic data. This method can be beneficial if trace shapes are consistent over long periods of time and signal in one or multiple orders is weak, which is often the case for STIS echelle data. Since STIS echelle gratings offer varying resolution and a variety of central wavelength settings, observations can vary greatly in their order separation on the detector, the angle/curvature of the orders incident on the detector, and S/N ratio. In order to further improve the flux recalibration of echelle secondary modes, a generic method is needed to derive trace locations of any given observation. In Section \ref{sec:obs}, we describe the calibration observations used to test and derive new secondary mode traces, in Section \ref{sec:method}, we describe the adopted trace-definition in detail, in Section \ref{sec:result}, we quantify overall improvements seen by implementing this new method, and in Section \ref{sec:conc} we summarize our new reference file deliveries (both pre- and post-SM4) for selected echelle modes and discuss how to perform future updates. 


\lhead{}
\rhead{}
\cfoot{\rm {\hspace{-1.9cm} Instrument Science Report STIS 2024-03 Page \thepage}}

\vspace{-0.3cm}
\ssection{Observations}\label{sec:obs}

Similar to STIS ISR 2012-01, all observations used for trace rederivation were taken from calibration program 11866 which observed the HST primary standard white dwarf (WD) G191-B2B in a variety of modes achieving a peak S/N ratio of 30 per pixel in E140M, E140H, and E230M and 20 per pixel for E230H. In regions where continuum flux is present, the minimum S/N ratio is 4 per pixel. A S/N ratio greater than 5 per pixel yields the best results for accurate trace definitions. The observations used for the trace redefinition in this ISR are listed in Table \ref{tab:obs}.

\begin{table}[!h] 
  \centering
    \caption{Summary of observations used to update echelle traces.}
    \def\arraystretch{1.25}
    \begin{tabular}{c c c c c c}
    \hline
    \hline
    Grating & Central Wavelength & Dataset & Observation Date & Exposure Time (s) \\
    \hline
    E140H & 1271 & obb001010 & 2009-11-30 & 696 \\
    E140H & 1307 & obb001090 & 2009-11-30 & 654 \\
    E140H & 1343 & obb005040 & 2009-12-01 & 3100 \\
    E140H & 1489 & obb0010a0 & 2009-11-30 & 1200 \\
    E230H & 2463 & obb053040 & 2010-01-06 & 670 \\
    E230H & 2713 & obb053050 & 2010-01-06 & 900 \\
    E230H & 2812 & obb0530a0 & 2010-01-06 & 1097 \\
    E230H & 2912 & obb053090 & 2010-01-06 & 1800 \\
    E230M & 2415 & obb004040 & 2009-11-29 & 280 \\
    \hline
    \end{tabular}
    \label{tab:obs} 
\end{table}

\ssection{Methods}\label{sec:method}

The STIS echelle secondary modes are diverse in their characteristics. In particular, WD standard star observations across different gratings and central wavelengths vary in their average S/N ratio, order spacing, angle of incidence, location of spectral features, and location of repeller wire shadows (present in E140H observations). These factors necessitate an adaptable approach to defining traces that can be implemented universally, without requiring tailored assistance for specific modes.

In the following sections we describe our method, which includes algorithms designed to first detect all orders incident on the detector, then fit the cross-dispersion profile of each order pixel-by-pixel across the detector, and finally apply a Gaussian process smoothing to these measurements resulting in a final trace for each order. These methods have been tested on a variety of STIS echelle modes and should be usable for refining the traces for other settings. 

\ssubsection{Order Identification and Cross-Dispersion Profile Fitting}\label{subsec:order}

\begin{figure}[!h]
  \centering
  \includegraphics[width=5.8in]{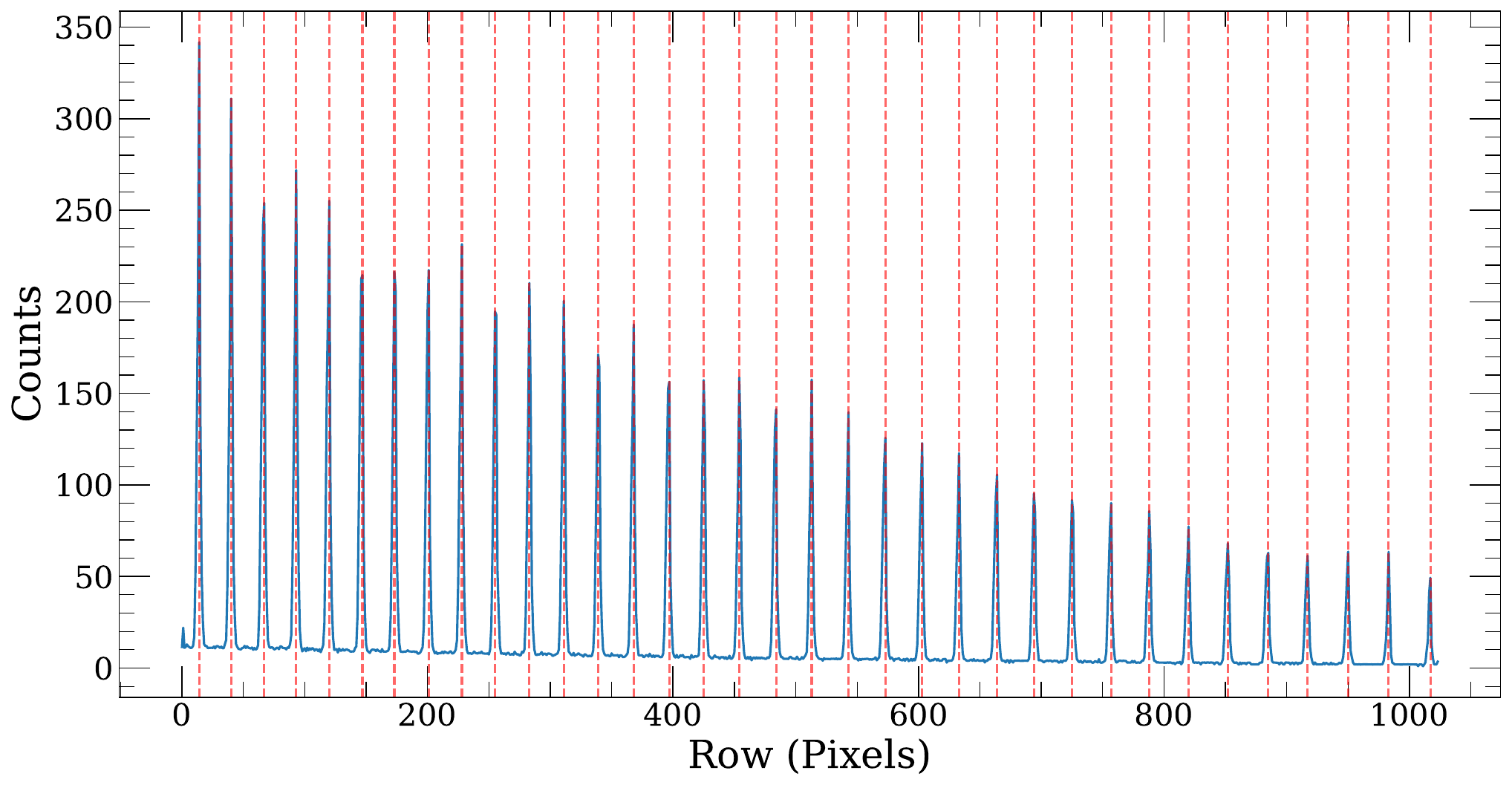}
    \caption{Example identification of echelle orders that are present on the detector for an observation taken with the E140H/1562 setting. The blue curve is a median crosscut of the central 50 columns of the detector, and the red-dashed lines are our measured fiducial positions of each order. These positions serve as a starting point for the defining regions of the detector where we measure the location of the cross-dispersion profile for each order. }
    \label{fig:order_identification}
\end{figure}

Each echelle order incident on the detector is identified by first taking a median crosscut using the central 50 columns of the detector, then applying a peak finding algorithm provided by the \code{scipy} python library. We find that a peak prominence threshold (defined by the difference between the peak counts and base level counts) equal to 10\% of the largest peak is necessary to correctly identify all orders present across all modes. Figure \ref{fig:order_identification} shows an example of this method using an observation of G191-B2B with E140H/1562. The blue curve is the median crosscut, and the red lines correspond to the row locations of each order near the central column of the detector. In this work, we only update existing traces in the previous reference files. Several observing modes have orders that fall near the edge of the detector, but are not extracted/flux calibrated. Since our method automatically locates these orders, future updates to include these missing orders can also use this trace derivation method. 

After identifying the orders incident on the detector, we iterate through each order fitting the cross-dispersion profile of the trace pixel-by-pixel in the dispersion direction. For most of the orders identified by our algorithm, straight-line traces have already been determined for the default pipeline extraction, and the existing EXTRLOCY array can be used to define the initial y positions for our Gaussian fits. For the other orders (often at the top and bottom edges), instead we use our measured peak locations to define the y-region. As done in STIS ISR 2007-03, we perform a 3-pixel boxcar smoothing on each column prior to the Gaussian fit in order to alleviate problems associated with undersampling of the cross-dispersion profile along the slit. We then fit a Gaussian plus constant background to each column in order to identify the y-location of the trace at every x pixel. This approach differs from some trace derivation methods that first smooth in the dispersion direction. Our approach allows for the better characterization of traces that vary on different length scales across the detector.

\begin{figure}
  \centering
  \includegraphics[width=5.5in]{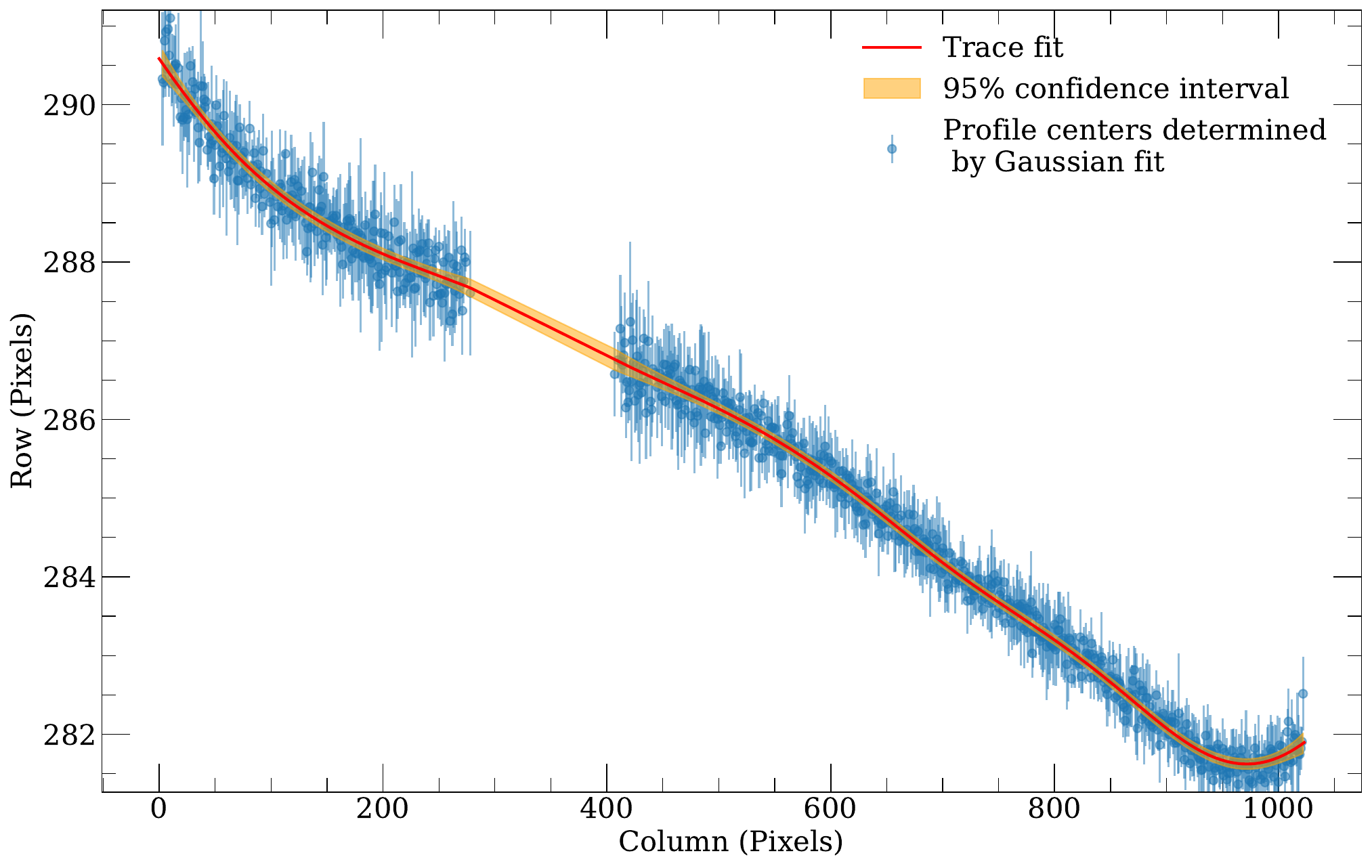}
  \includegraphics[width=5.5in]{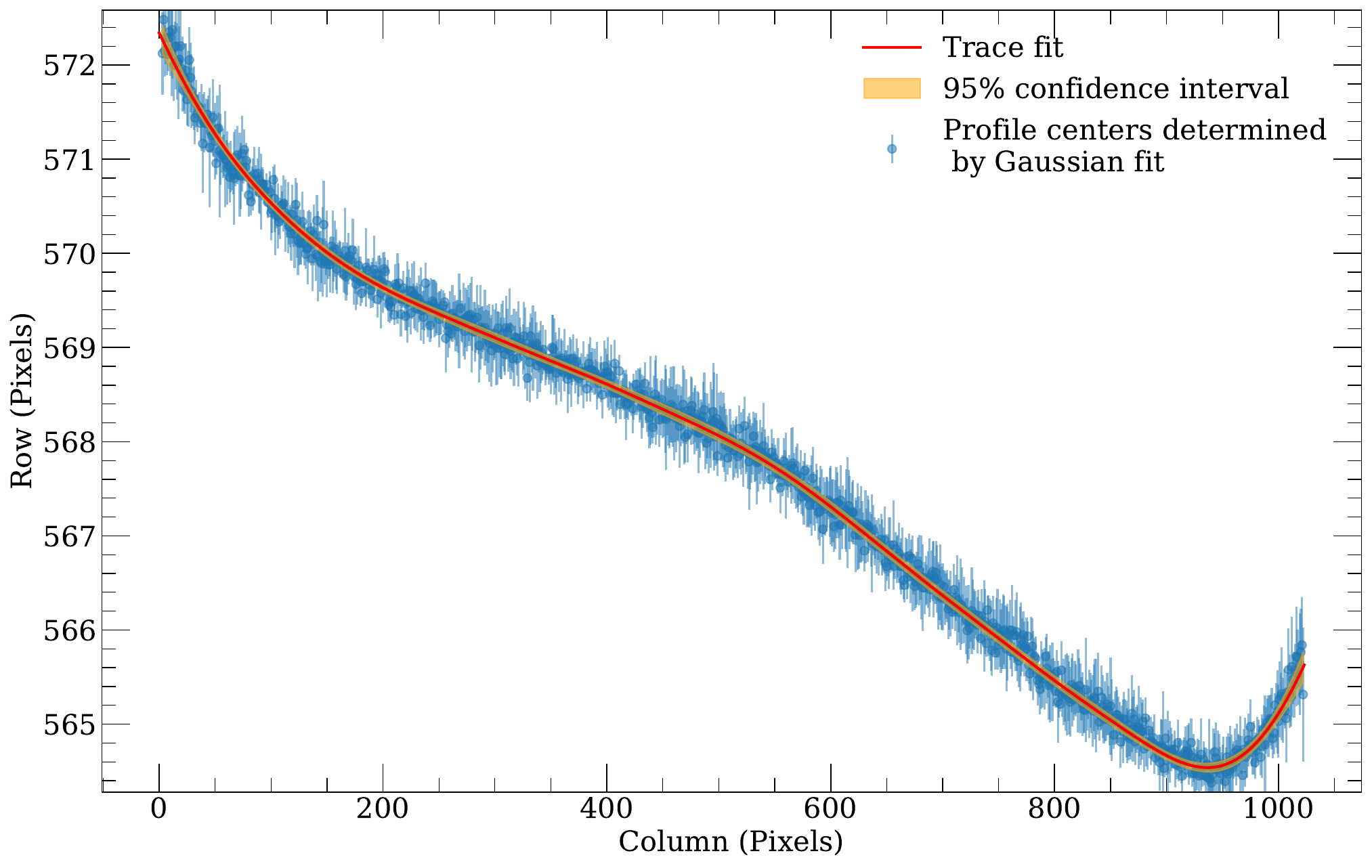}
    \caption{Example trace fits using orders 346 (top) and 331 (bottom) from E140H/1271. The blue curve is the measured pixel-by-pixel cross-dispersion profile location, and the shaded blue region is the associated uncertainty. The red curve is our final trace definition resulting from a Gaussian process fit to the data, and the orange curve is the 95\% confidence interval of the fit. This method smoothly interpolates over regions of the detector with low S/N (Lyman-$\alpha$ in the top panel), and reproduces the broad curvature of the order. This allows for proper centering of the cross-dispersion profile in the extraction region over the full wavelength extent of each order.}
    \label{fig:gaus_fit}
\end{figure}

To ensure only high S/N data contribute to the re-definition of the traces, we exclude fits where the peak of the cross-dispersion profile is less than 5 times the measured background level. The choice to model the cross-dispersion profile as a simple Gaussian may have led to underestimated uncertainties in the profile centers. By iteratively increasing these uncertainties, we have found that a scaling factor of 6 reflects the observed scatter in these measurements across a wide variety of gratings and central wavelengths settings. The subsequent Gaussian-process trace smoothing better represents the broad curvature of each order in traces that were previously overfit. In Figure \ref{fig:gaus_fit} we show example trace measurements and 1-$\sigma$ uncertainties for E140H/1271, orders 346 and 331 (blue curves; top and bottom, respectively).

\ssubsection{Gaussian-Process Modeling of Echelle Traces}\label{subsec:gaus}

The final step required for defining the trace of an order is smoothing of the noisy pixel-by-pixel measurements of the trace location in the dispersion direction. Spline fits have been used to define traces for the first-order gratings (see STIS ISR 2007-03) but the complexity of echelle trace shapes requires a different solution. Figure \ref{fig:gaus_fit} shows two different example traces  measured from the cross-dispersion profile fitting described in Section \ref{subsec:order} (blue points). In some orders the trace is relatively linear, while in others the trace can turn up at the edges of the detector requiring a different length scale for the trace definition. Furthermore, regions with very low S/N in the calibration data, often caused by strong absorption features or repeller wire shadows, need to be carefully considered when defining traces. The scale of trace variability near the detector edges and  the size of low S/N regions vary greatly across different gratings and cenwaves. Using a spline fitting approach would require defining order-specific spline knot locations, spacing intervals, and low S/N region interpolation techniques, requiring careful and time-consuming testing. 

Instead, we have developed a different fitting method that uses Gaussian process regression techniques to provide a universal solution. In our case, we do not know the exact functional form that represents how the trace shape changes across the detector. Since a Gaussian Process aims to model the distribution of possible functions that could explain the data, this method is an appropriate choice for modeling echelle traces. The distribution of functions is represented by the mean and covariance function known as the kernel function. During the training phase, the Gaussian Process model estimates the parameters of the kernel (or covariance) function based on the provided training data (i.e., blue points in Figure \ref{fig:gaus_fit}). 

In the context of 1D interpolation of a spectral trace, each order is treated separately, and the training data are the measured cross-dispersion profile centers of the trace. The kernel function informs the model on the similarity of all pairs of data points. For example, in a spectral trace, we expect a high covariance of adjacent pixels, and low covariance for pixels on opposite edges of the detector. The simplest choice of kernel is known as the Radial Basis Function (RBF) which simply expresses similarity between pixels via their Euclidean distance. A Matern kernel is a generalization of the RBF kernel that includes an additional smoothness parameter. In testing, we observed that the Matern kernel yielded better results in characterizing the traces of spectral orders with stronger upturns near the edges of the detector. This improvement is likely due to the Matern kernel having greater flexibility in describing pixel covariance on varying length scales within a single order. 

Prior to Gaussian process training we first rescale the input data (blue curves in Figure \ref{fig:gaus_fit}). We have chosen a  Matern kernel\footnote{\url{https://scikit-learn.org/stable/modules/generated/sklearn.gaussian_process.kernels.Matern}} with an initial length scale of 10, smoothness parameter of 2.5, and length scale bounds between 3 and 20 which is used across all orders, cenwaves, and gratings. We then use our trained Gaussian process model to predict the location of the trace at every column of the detector (red curves in Figure \ref{fig:gaus_fit}).  

\begin{figure}
  \includegraphics[width=2.8in]{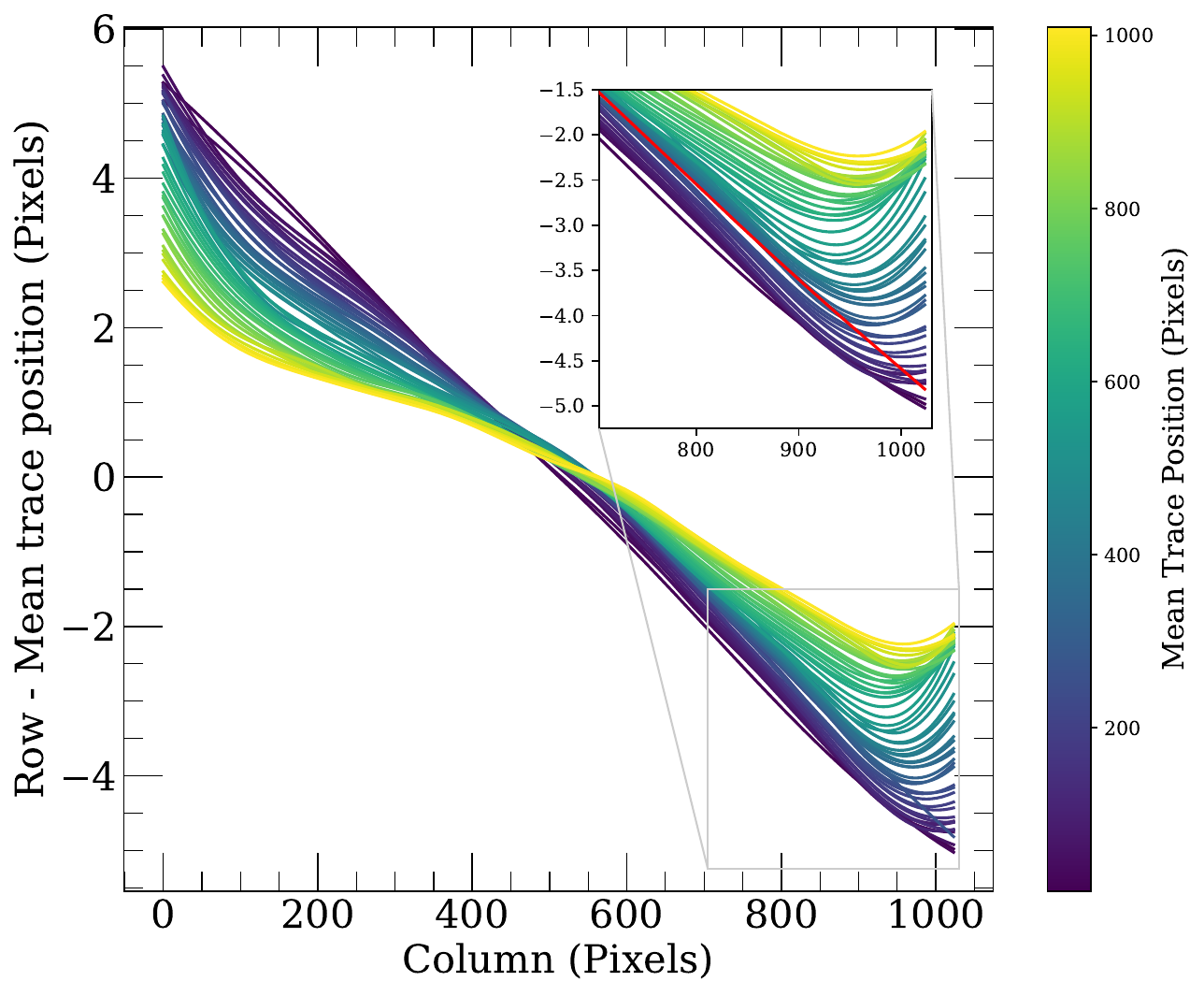}
  \includegraphics[width=2.8in]{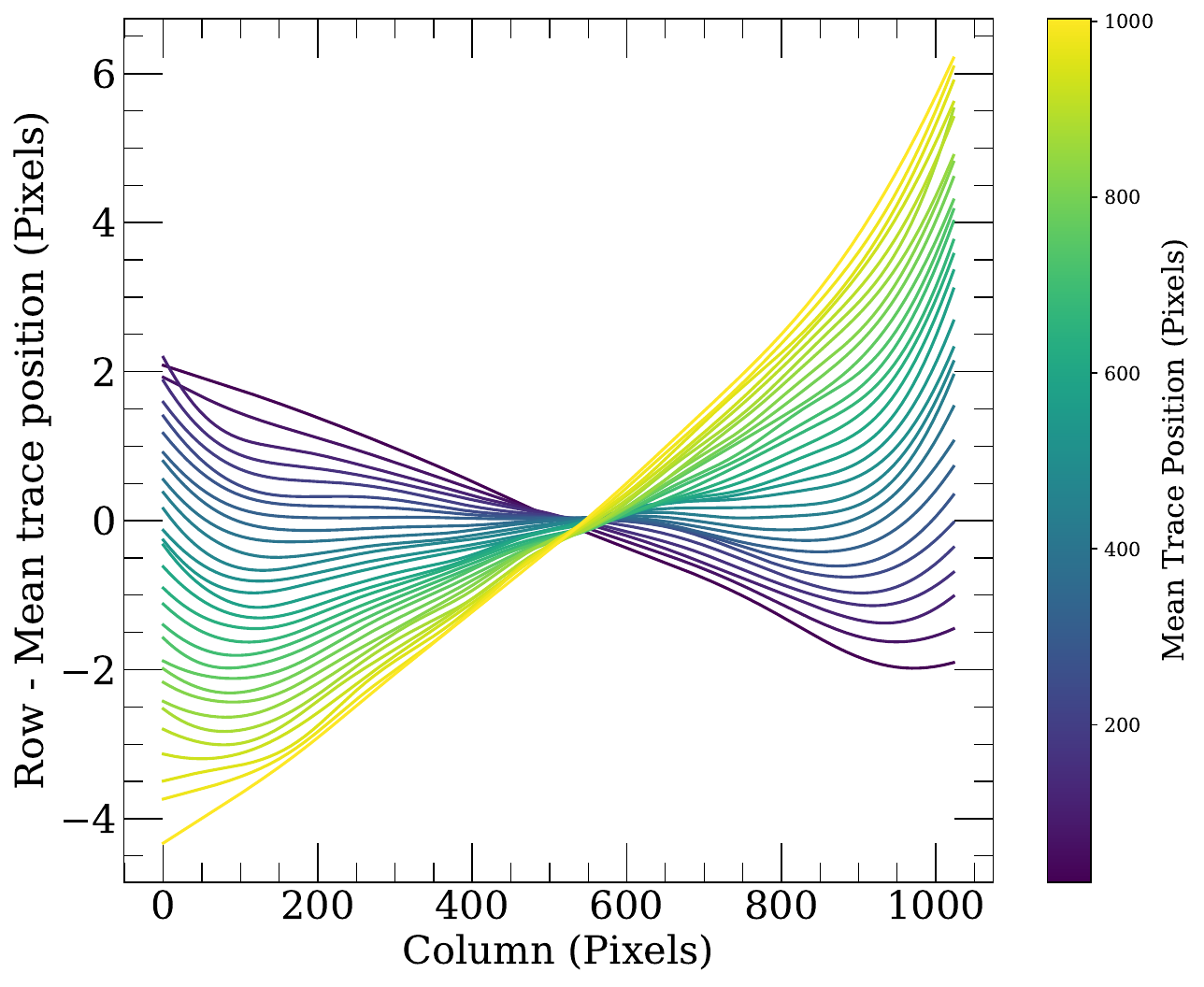}
    \caption{Variation of spectral trace shapes along the cross-dispersion direction of the detector. E140H/1271 is displayed on the left and E230M/2415 is displayed on the right. The mean trace position is subtracted from the spectral trace of each order so shapes can be easily compared. Regions of low S/N where the cross-dispersion profile cannot be located can lead to extrapolation of the spectral trace at the edges of the detector. An example of this effect is shown in the trace of order 347 (red curve) in the inset of the left panel. While adjacent orders show a slight upturn at the edge of the trace, order 347 is instead extrapolated. }
    \label{fig:trace_variation}
\end{figure}

In Figure \ref{fig:trace_variation}, we show how the traces defined using this method vary for both E140H/1271 (left) and E230M/2415 (right) across the detector. Generally the shape of the trace changes smoothly from order to order. For all grating/cenwaves the slope of the trace gradually changes, rotating the trace counter-clockwise when going from a low to high numbered row.

Currently, adjacent orders are not used to inform the trace fitting of any individual order which can lead to some minor issues at low S/N. When S/N is less than 5 near the edge of the detector, the trace must typically be extrapolated. The resulting shape is often straighter than adjacent orders might suggest. For example, this occurs in two locations in E140H/1271. First, the bluest edges of the two bluest orders (360 and 361), and second, order 347 strong Lyman-$\alpha$ absorption lands on the red edge of the trace. We show how our Gaussian-process method extrapolates this region in \autoref{fig:improve2} in the Appendix. A region of low-S/N also likely affects the trace shape of the bluest order of E230M/2415. While these instances may not be ideal, we note that the resulting extrapolation is still better than the previous straight-line fits. 

Gaussian process regression has several major advantages. First, this method not only provides the best fitting trace, but also estimates the associated uncertainties (orange region). Second, this method can naturally predict the trace location where there are large gaps in data due to low S/N. Lastly, the final traces are derived using the best fitting length scale for every order, automatically accounting for variability in the trace near the edges of certain orders. After iterating through every identified order and defining a final trace for each, we apply these changes to the relevant SPTRCTAB reference files. Updates are made by modifying the A2DISP keyword, which consists of offsets in pixels in the AXIS2 direction relative to A2CENTER. While we have chosen to update only a selection of secondary echelle modes of interest in the latest flux recalibration release, we expect that this procedure can be used in the future to redefine traces for the rest of the secondary modes. 

\ssection{Results and Accuracies}\label{sec:result}

In this section, we compare our newly defined traces to past methods and show the improvement in S/N achieved for a variety of STIS echelle secondary modes. We discuss how updating secondary traces is a necessary step to further improve the STIS flux recalibration effort, and show how these updates affect newly-derived sensitivity functions. 

\ssubsection{Comparison to Primary Modes and other Standard Stars}\label{subsec:primary}

\begin{figure}[htb!]
  \centering
  \includegraphics[width=5.0in]{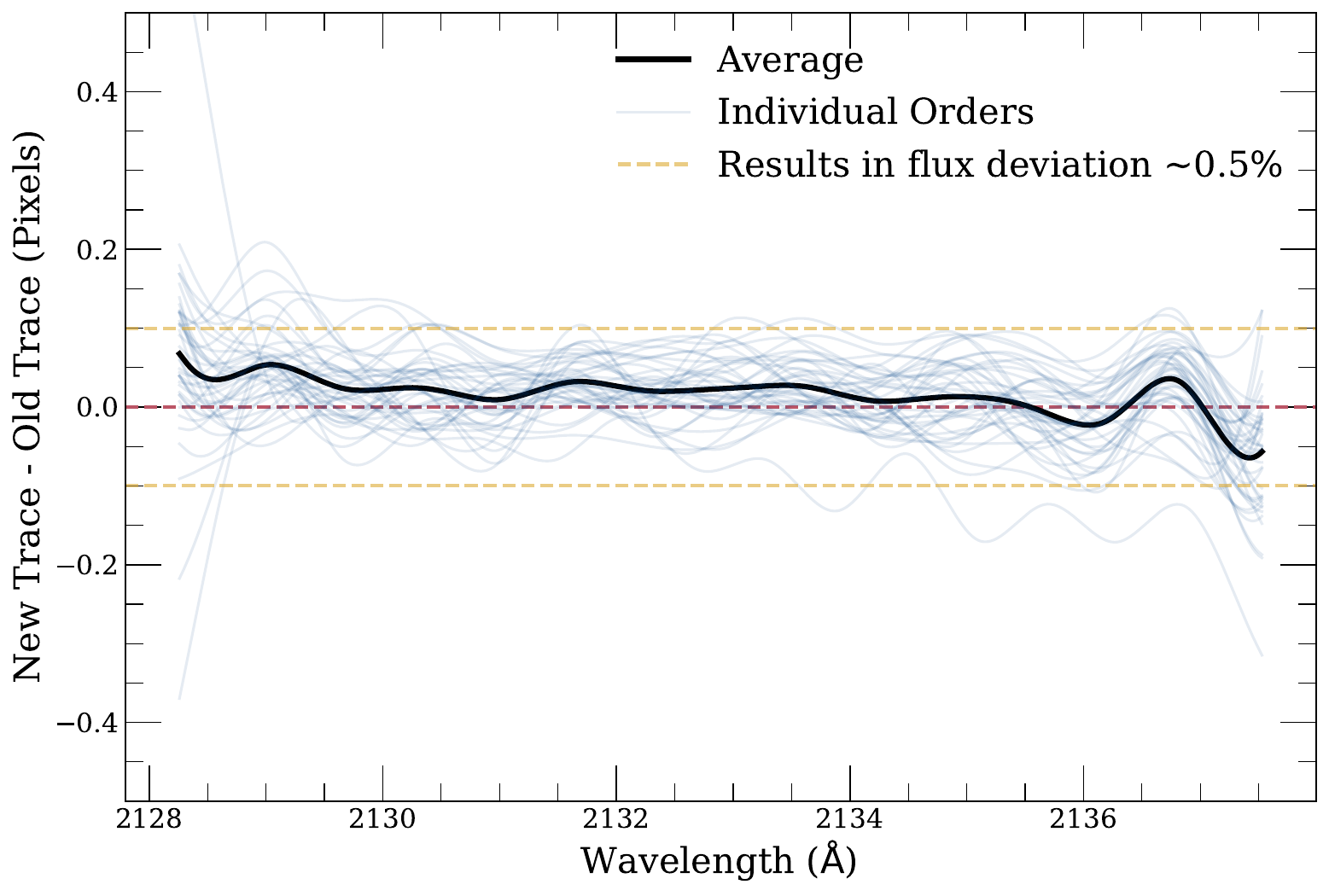}
    \caption{Difference in pixels as a function of wavelength between current primary mode (E230H/2263) traces to the trace we define for the same orders using our new method (blue curves). The black curve shows the average difference over all orders and we find typical differences on the order of 0.05-0.1 pixels which propagate to flux differences that are $< 1\%$. The largest differences occur at the order edges where S/N is very low and the trace positions need to be extrapolated.}
    \label{fig:primary}
\end{figure}

\begin{figure}[htb!]
  \centering
  \includegraphics[width=5.0in]{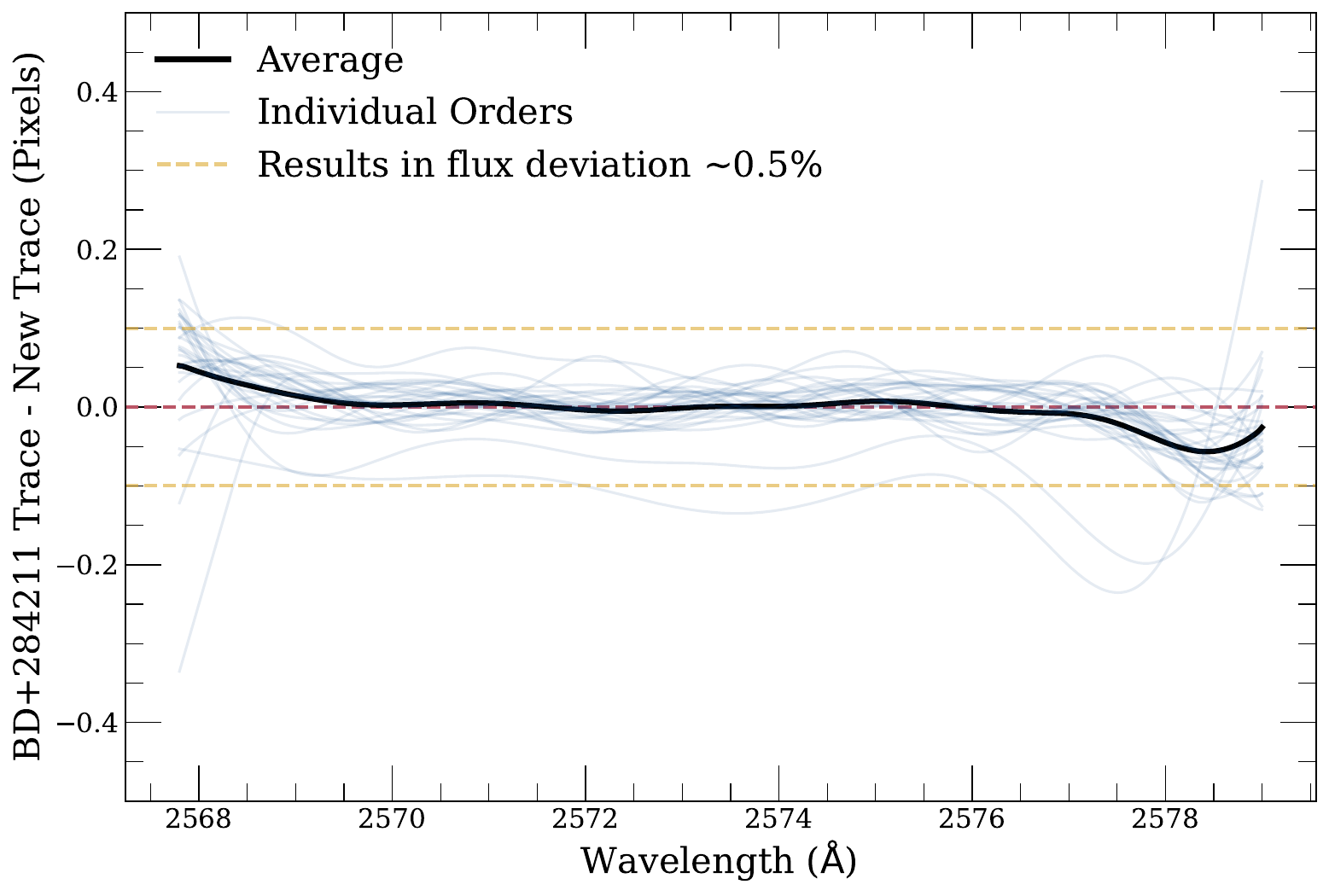}
    \caption{Difference in pixels as a function of wavelength between traces derived using E230H/2713 observations of a different standard star (BD+284211) and G191-B2B (this work). The black curve shows the average difference over all orders and we find typical differences on the order of $\sim 0.05$ pixels.}
    \label{fig:bd28}
\end{figure}

As shown in \autoref{fig:trace} (top panel), primary echelle modes already have curved traces. To assess the accuracy of our new trace derivation method, we have also derived new traces for the primary mode E230H/2263. In Figure \ref{fig:primary}, we compare our newly-derived traces to the current traces defined for the E230H/2263. We display this as the difference between the new trace and previous trace in pixels for each order (blue curves) as a function of wavelength. The black curve is the average difference. The dashed-yellow curves correspond to deviations of 0.1 pixels, which translate to deviations in flux of $\sim 0.5\%$. The differences in traces mostly fall within $\pm0.1$ pixels. Across all orders for E230H/2263, we measure an average deviation of $0.02\pm0.02$ pixels indicating that our new method performs similarly to the one used for primary echelle modes. The largest deviations occur at the order edges and in the most extreme cases, can be as large as $\pm0.4$ pixels. However, we note that these large differences only occur when the flux near the edge of the detector is close to zero, and the trace needs to be extrapolated. Thus, we can confidently apply this new method to the secondary echelle modes whose traces have only previously been defined with linear fits.

Since the calibration program PID 11866 observed the standard star G191-B2B in all STIS echelle modes, we have opted to use these data to define new traces. However, one might be concerned that the choice of a single standard star may lead to biased results. To test this, in Figure \ref{fig:bd28}, we compare traces derived using our Gaussian process method for observations of two different standard stars, G191-B2B and BD+284211. Given the abundance of secondary mode echelle calibration data for G191-B2B we have opted to use this standard star for all SPTRCTAB updates. With an average deviation of $-0.002\pm0.02$ among all E230H/2713 orders it is clear that this method is relatively insensitive to the calibrating star. Similar to the primary mode trace comparisons, the largest deviations occur in regions of low S/N.

\ssubsection{Improvements Over Previous Secondary Traces}\label{subsec:improve}

\begin{figure}[!h]
  \includegraphics[width=5.8in]{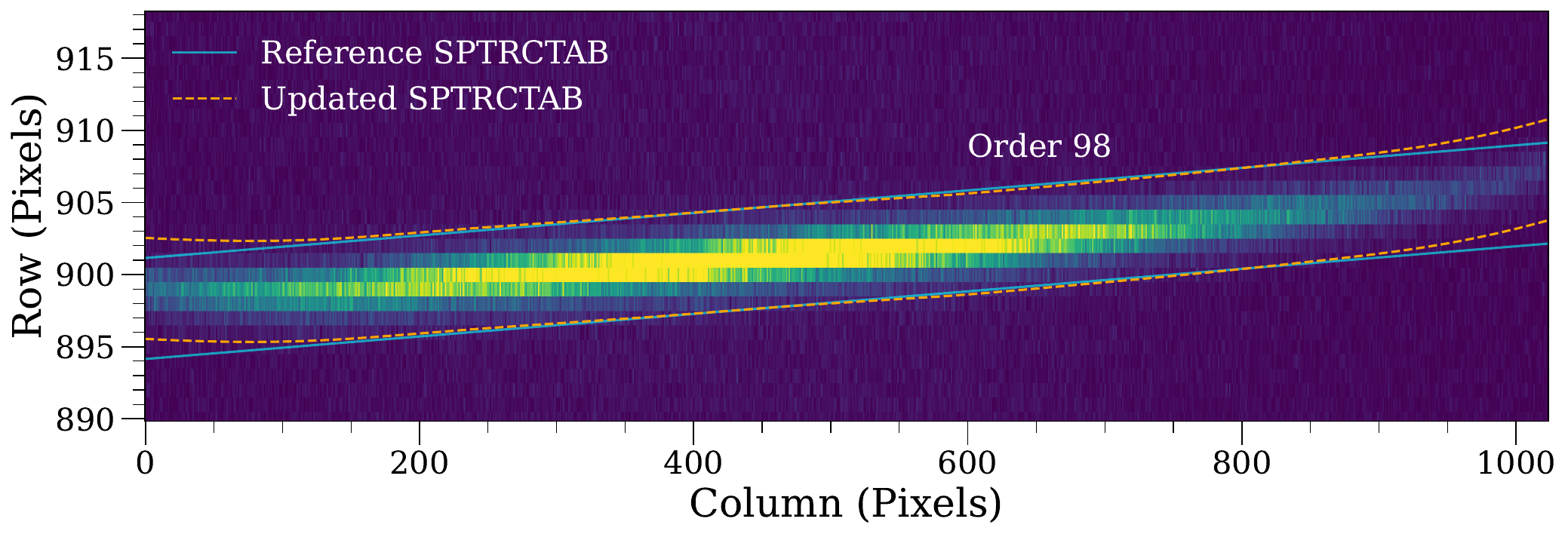}
  \includegraphics[width=2.9in]{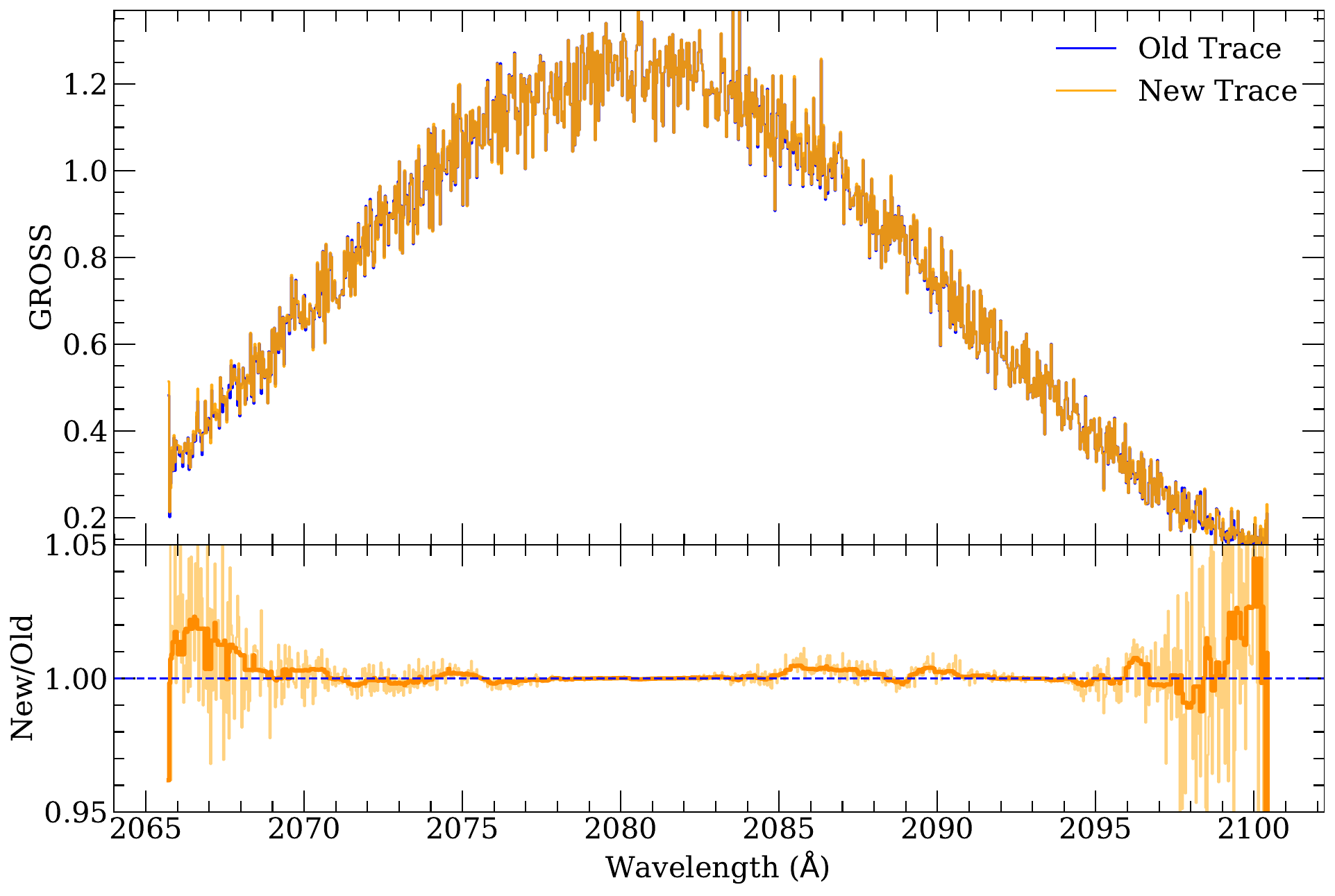}
  \includegraphics[width=2.9in]{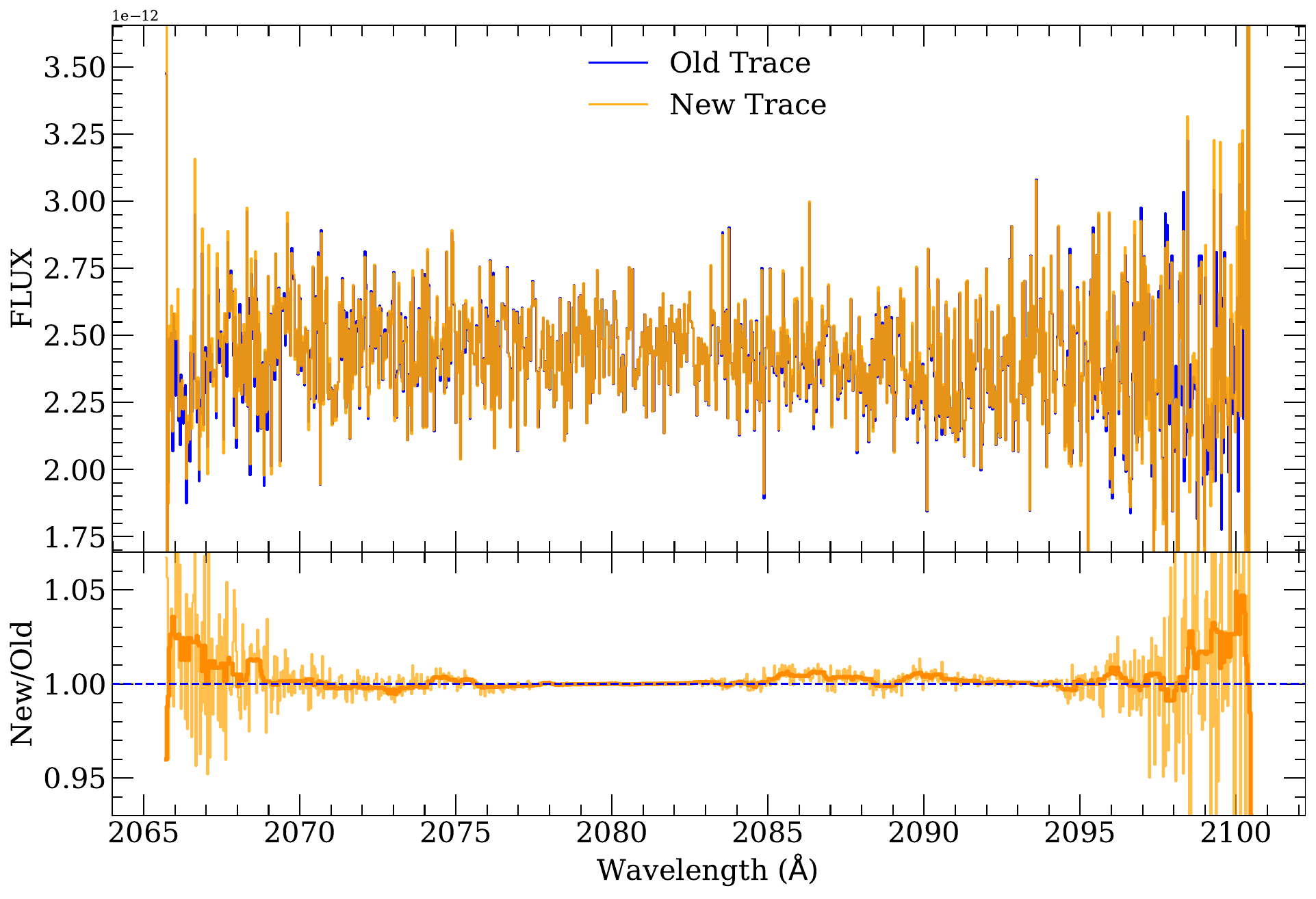}
    \caption{Example improvements (E230M/2415, order 98) in extraction region (top), gross counts (bottom-left) and flux (bottom-right), when reducing data using the previous straight-line secondary traces (blue curves) and our new Gaussian process-defined traces (orange curves). The dark orange curves are 15-pixel median filter version of the underlying ratios. At the edges of the detector, improvements in the trace location of $\sim 0.5 - 2$ pixels lead to the recovery of $\sim 2 - 4\%$ more flux in these regions.}
    \label{fig:improve}
\end{figure}

We have applied our new trace-derivation method to a variety of STIS echelle secondary modes using observations from PID 11866. Overall, we find that this new method provides traces that better represent both the broad overall curvature and the smaller scale curvature near the ends. A representative example of the impact of these changes is presented in \autoref{fig:improve}. The top panel shows a 2D section centered on order 98 (E230M/2415) of an observation of G 191-B2B. The y-axis has been stretched to highlight differences in the spatial direction. The blue lines show the original extraction region resulting from the straight line trace, and the orange lines show the extraction region resulting from our new trace-derivation method. The largest differences between methods occur near the edges where our new method deviates by nearly 2 pixels from the original trace.  

The impact of these differences on the resulting gross counts and flux is shown in the bottom two panels of \autoref{fig:improve} (left and right, respectively). As expected the largest differences occur near the edges. Newly-derived traces tend to recover roughly 2-4\% more of the flux in these regions, and the impact on S/N is very similar. These improvements are observed across a large variety of gratings and central wavelength settings. In particular, for E140H/1271, E230M/2415, and E230H/2713, we find that 33\%, 55\%, and 62\% of the orders achieve improvements $>4\%$ in flux, respectively. We provide additional comparisons in the Appendix so readers can get a better idea of the general changes that we observe with this new trace-derivation method.

\ssubsection{Redefined Throughputs}\label{subsec:sens}

\begin{figure}[htb!]
  \centering
  \includegraphics[width=3.2in]{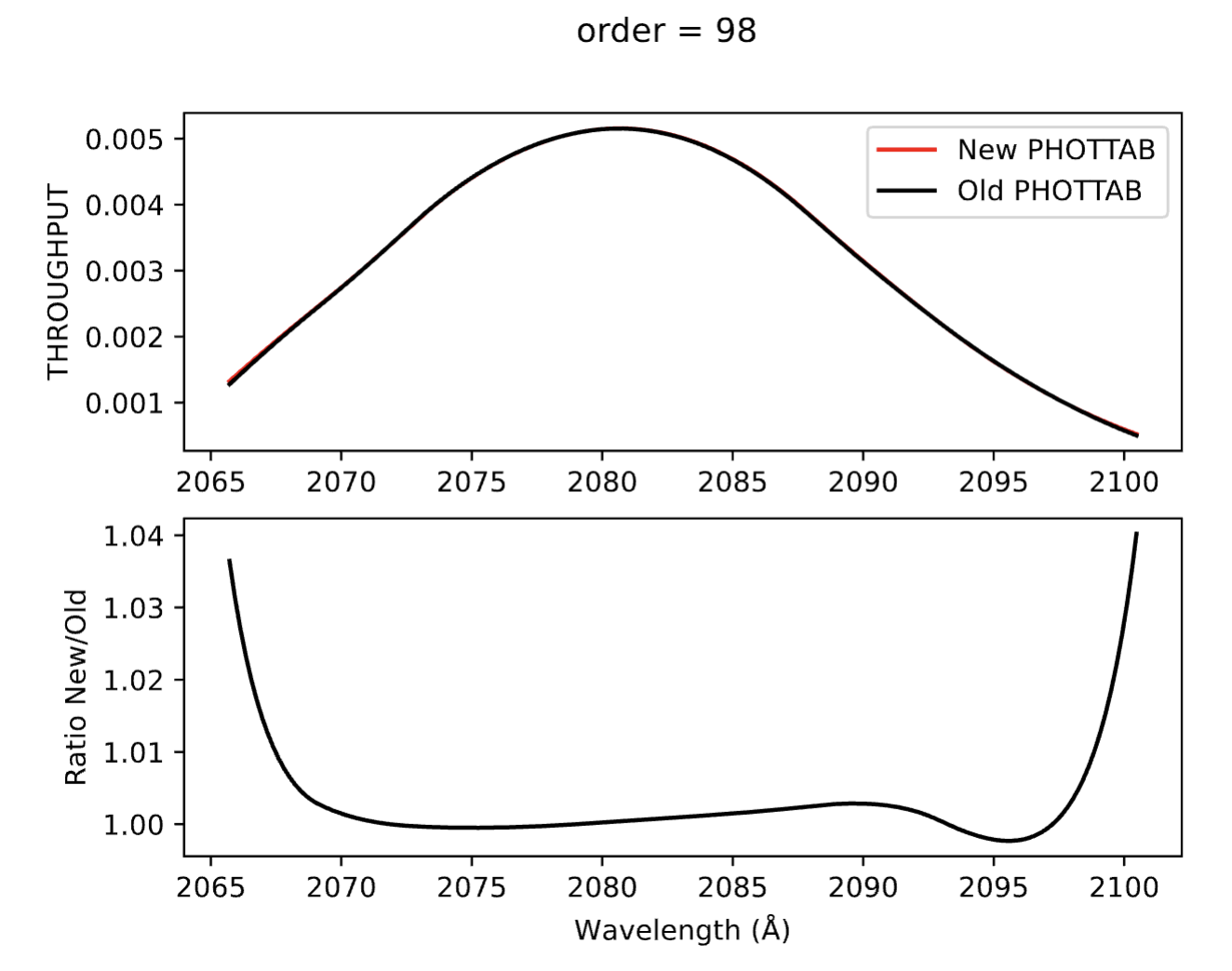}
    \caption{Comparison of the throughput for E230M/2415 order 98 derived using the original straight line trace definition (black) and the throughput derived using our new trace definition method (red). The ratio of the two curves is shown in the bottom panel. For most of the wavelength range, differences in throughput are $<1\%$. Differences up to $\sim 4\%$ are present near the order edges.}
    \label{fig:sens}
\end{figure}

For accurate flux calibration, these new traces must be used to rederive throughputs for each order of each echelle mode. In \autoref{fig:sens}, we show the updated throughput (red curve) for order 98 of E230M/2415 (same order as \autoref{fig:improve}) compared to the old throughput derived from the straight-line trace (black curve). We see that recovery of the flux near the edges of the detector has resulted in a corresponding increase in the throughput at the same wavelengths. This change is on the 2-4\% level as expected from the change in flux. 

\vspace{-0.8cm}
\ssection{Conclusions}\label{sec:conc}

We have developed a new method for the derivation of STIS echelle mode traces from calibration data. This method has primarily been designed to provide replacement traces for secondary echelle modes where the original traces were defined with straight line fits. New traces are derived using Gaussian process interpolation and this method can be applied universally without the need for echelle mode-specific changes, facilitating future updates if they are needed. These traces better-represent the overall curvature of each order, and provide improvements to flux throughput of about 1-4\% with the largest improvements typically near the detector edges. Both pre-SM4 and post-SM4 reference files have been updated with these new traces. Future work should investigate the impact of focus and aperture choice on spectral trace definitions. Additionally, the fitting method used in this ISR could be further improved by extending the Gaussian process interpolation into two-dimensions. This would allow adjacent higher S/N spectral orders to inform the shape of orders with regions of low S/N.

We have prioritized specific echelle modes based on historical usage for a first round of updates. These modes are: E140H/1271, E140H/1307, E140H/1343, E140H/1489, E230M/2415, E230H/2463, E230H/2713, E230H/2812, and E230H/2912. For two of these modes (E230M/2415, and E230H/2713), the STIS team has also delivered updated sensitivities and blaze shift coefficients as described in the January 2023 and July 2023 STAN articles.


\vspace{-0.5cm}
\ssectionstar{References}\label{sec:References}
\vspace{-0.3cm}

\noindent
Bohlin, R., Hubeny, I., \& Rauch, T. 2020, AJ, 160, 21\\
Bostroem, K. A., Aloisi, A., Bohlin, R., Hodge, P., \& Proffitt, C. 2012, STIS Instrument Science Report 2012-01\\
Medallon, S., Welty, D., et al. 2023 “STIS Instrument Handbook (IHB),” Version 22.0, (Baltimore: STScI)\\

\vspace{-0.3cm}
\ssectionstar{Appendix A}\label{sec:Appendix}
\vspace{-0.3cm}

\begin{figure}[!h]
  \includegraphics[width=5.8in]{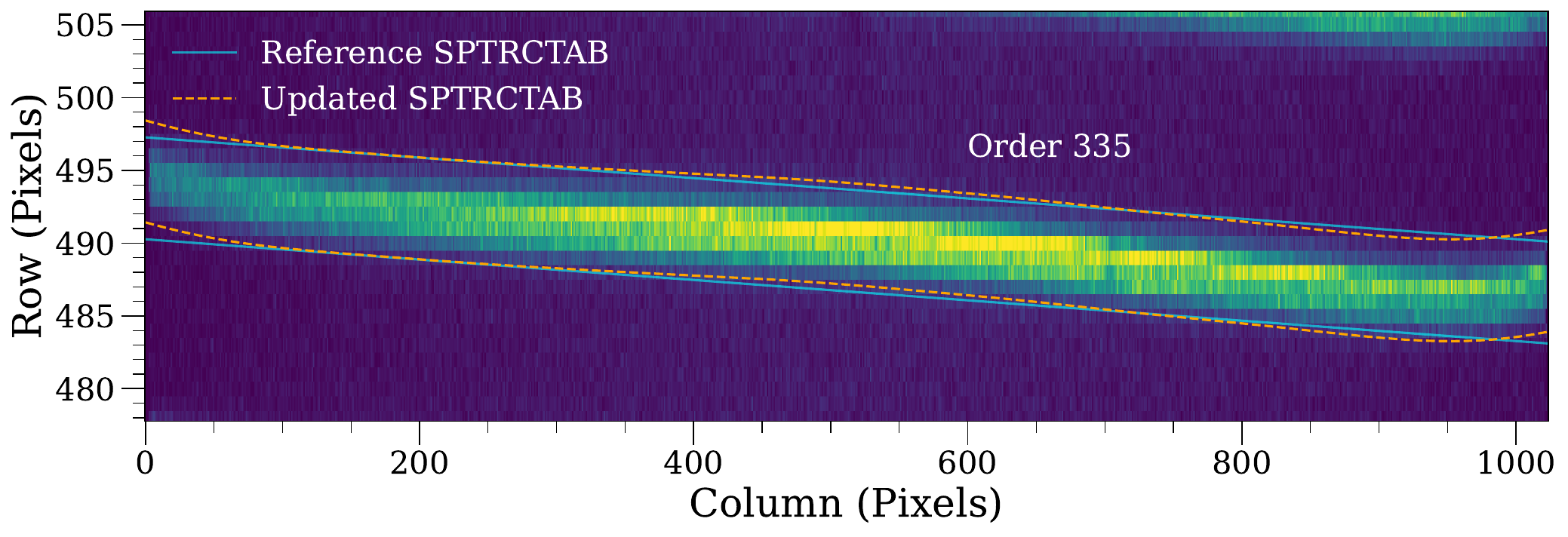}
  \includegraphics[width=2.9in]{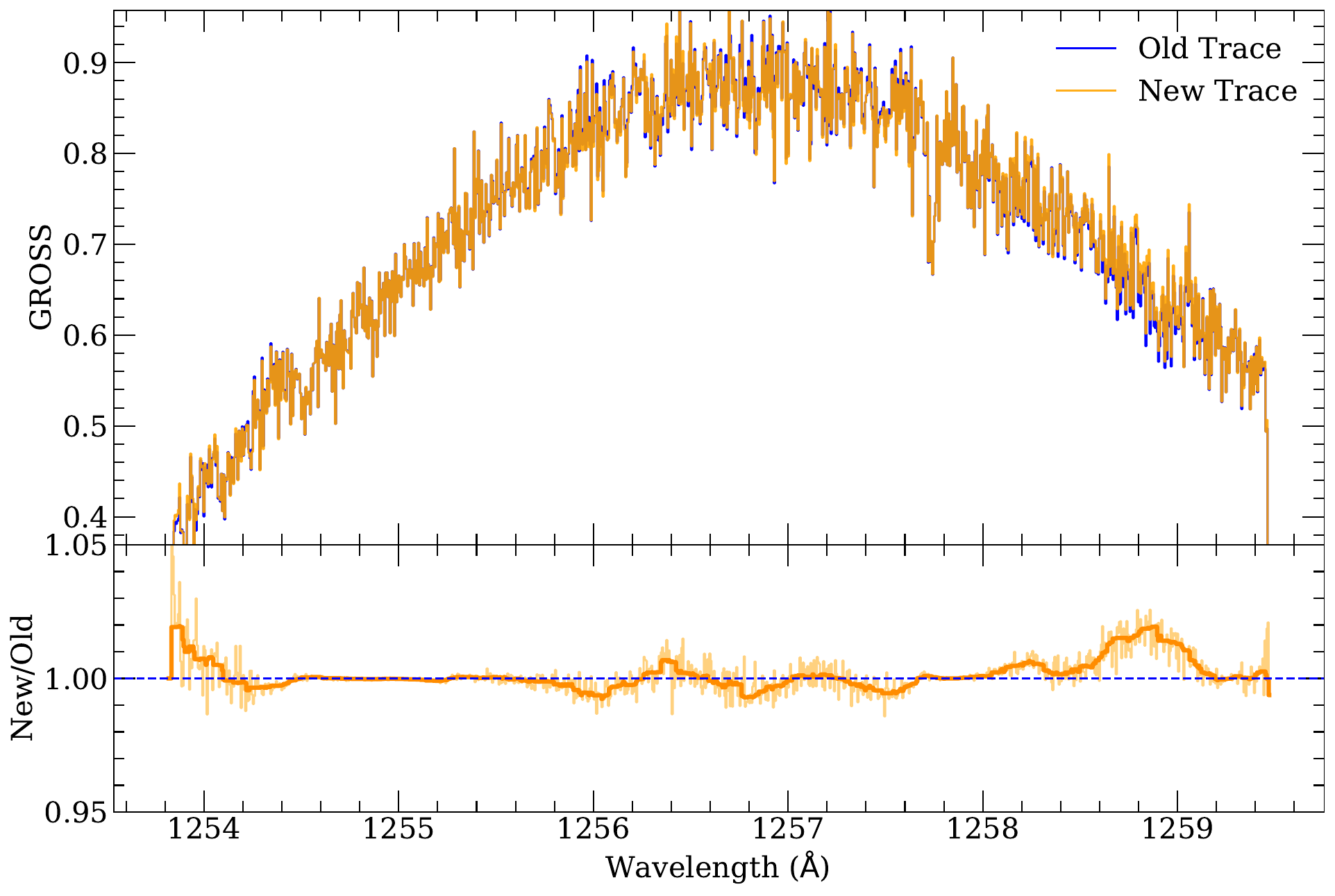}
  \includegraphics[width=2.9in]{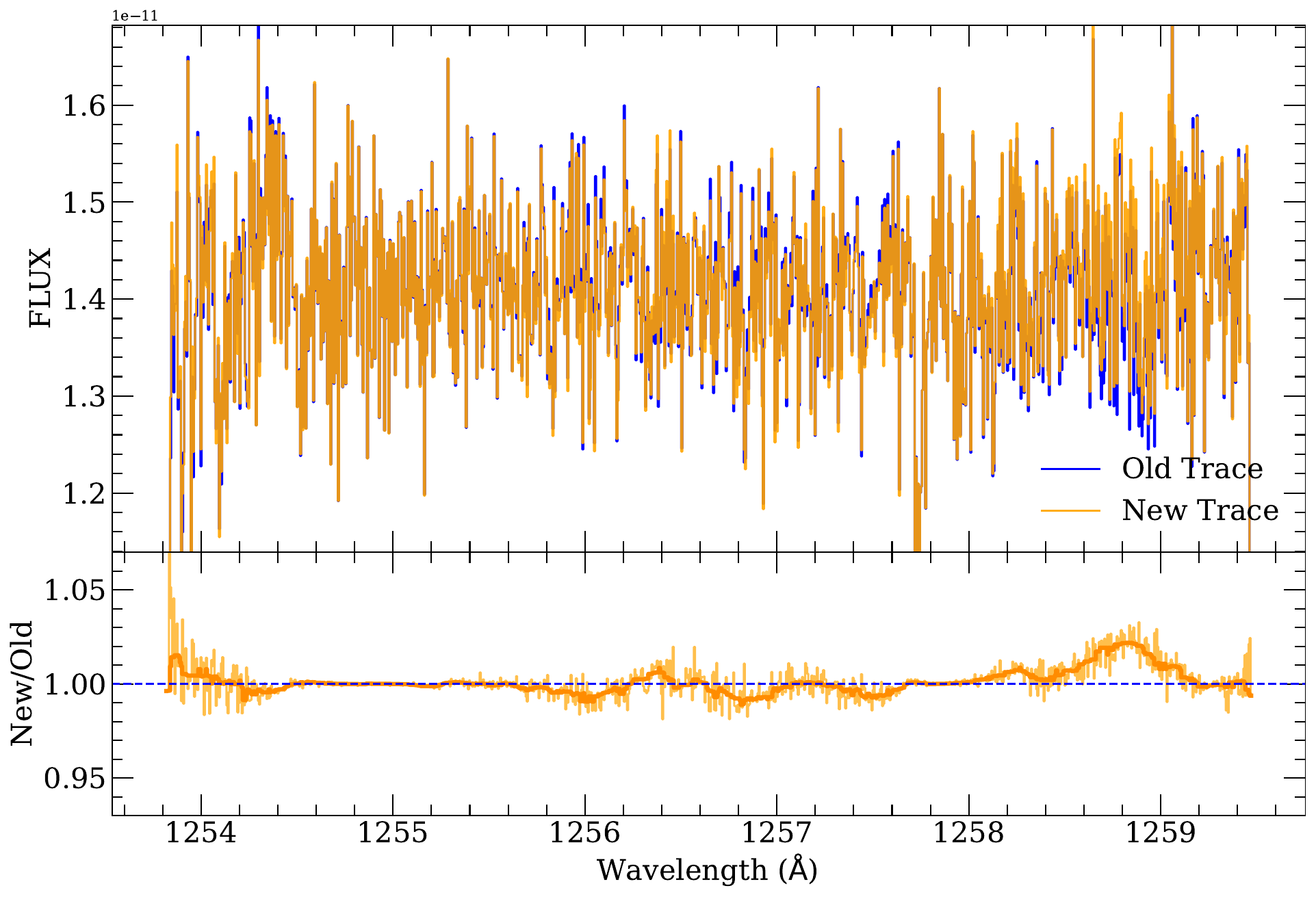}
    \caption{Example improvements (E140H/1271 order 335) in extraction region (top), gross counts (bottom-left) and flux (bottom-right) when reducing data using the previous straight-line secondary traces (blue curves) and our new Gaussian process-defined traces (orange curves).}
    \label{fig:improve2}
\end{figure}

\begin{figure}[!h]
  \begin{center}
   \includegraphics[width=4.8in]{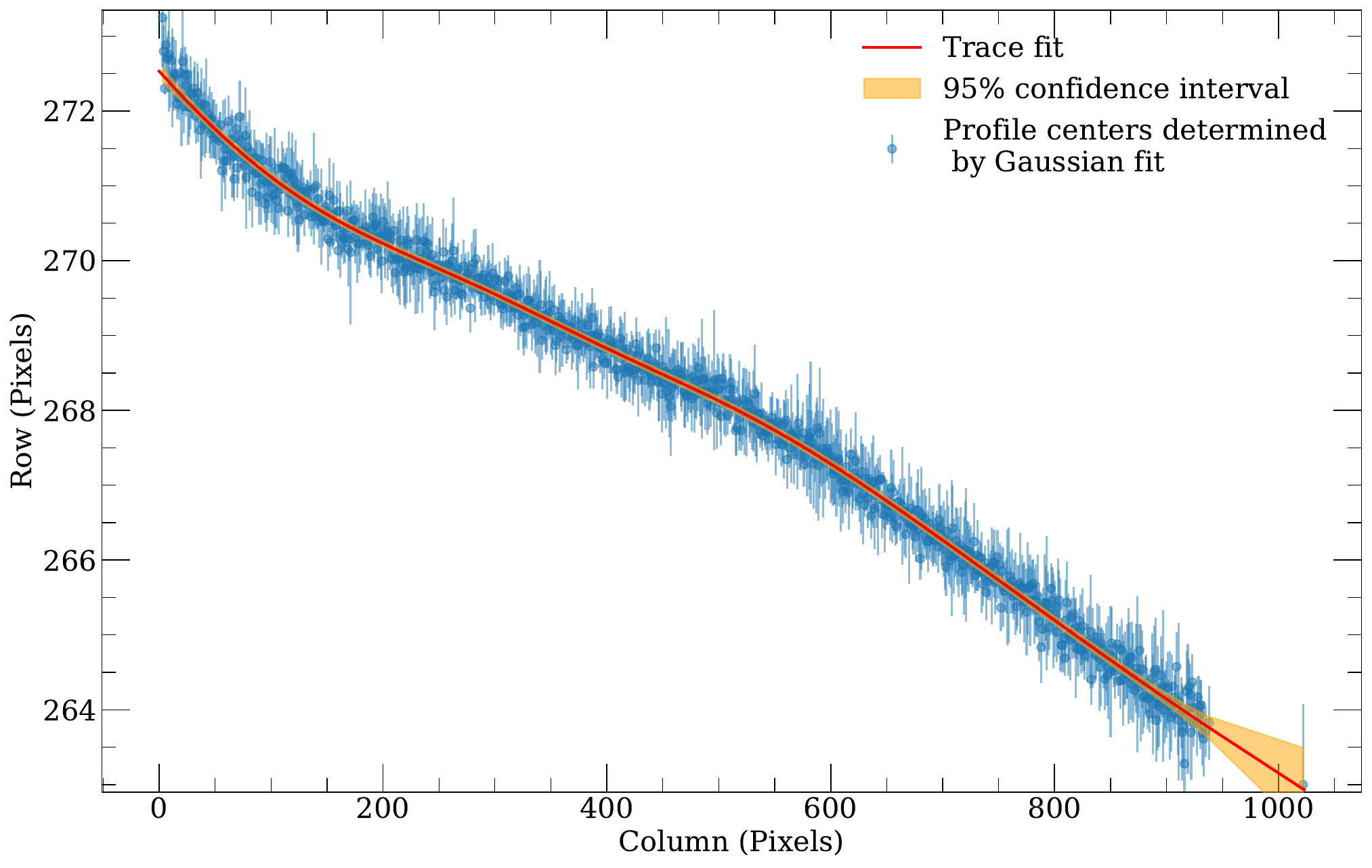}
   \end{center}
  \includegraphics[width=5.8in]
  {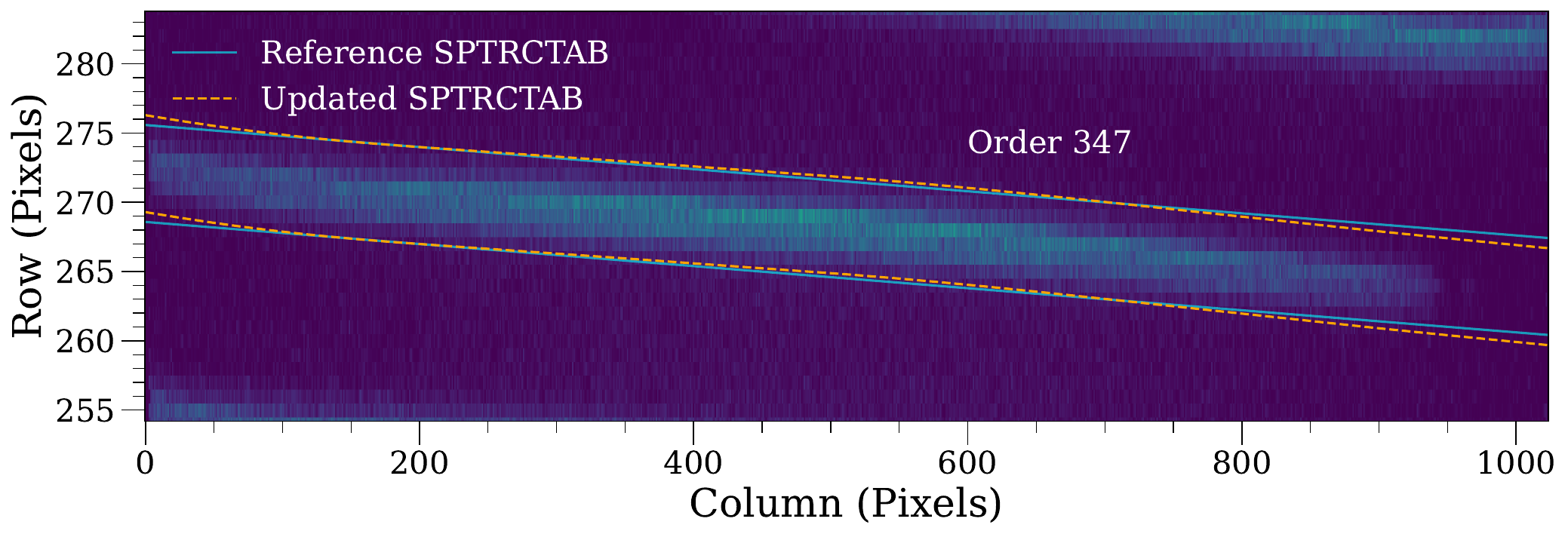}
  \includegraphics[width=2.9in]{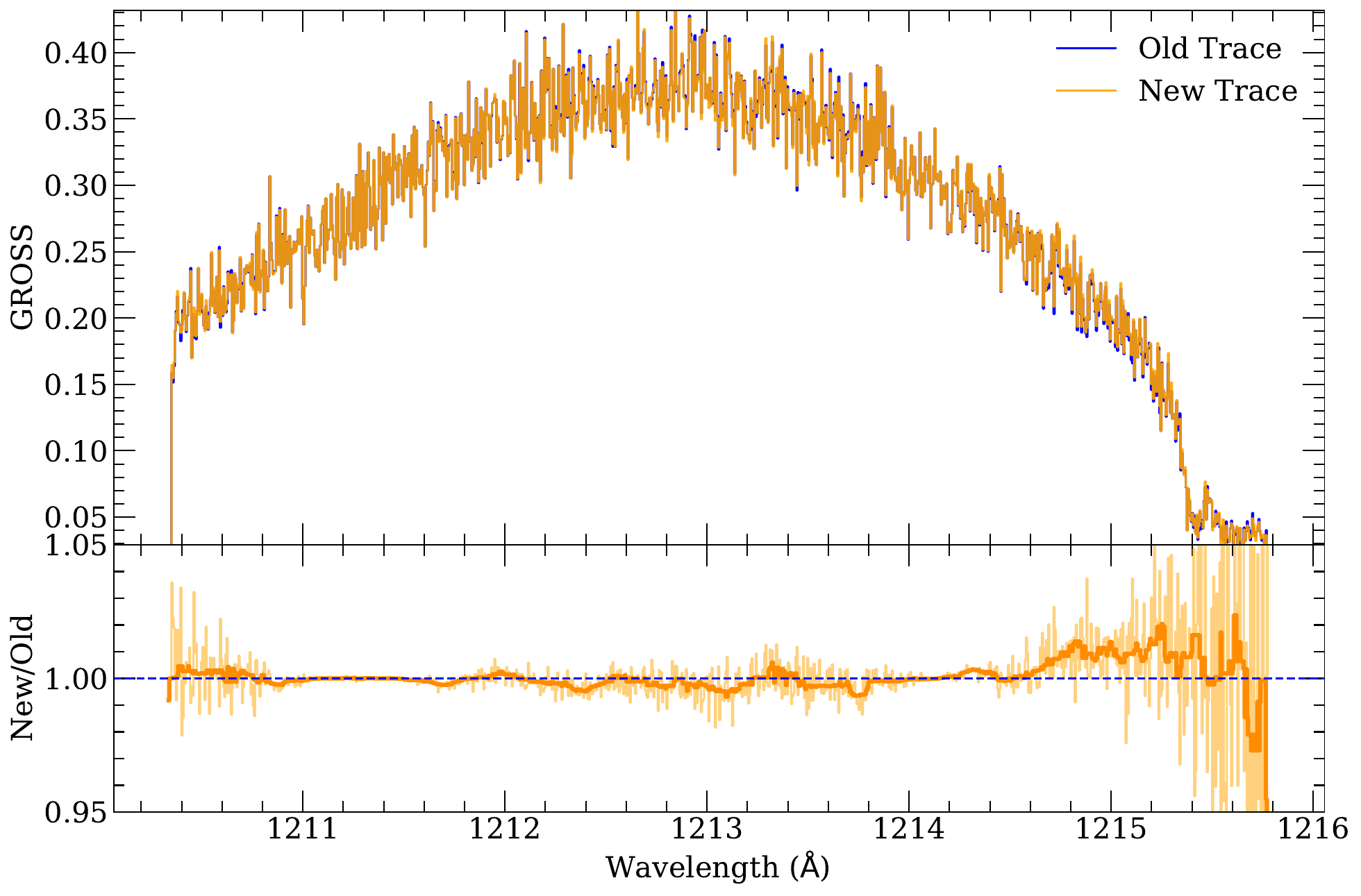}
  \includegraphics[width=2.9in]{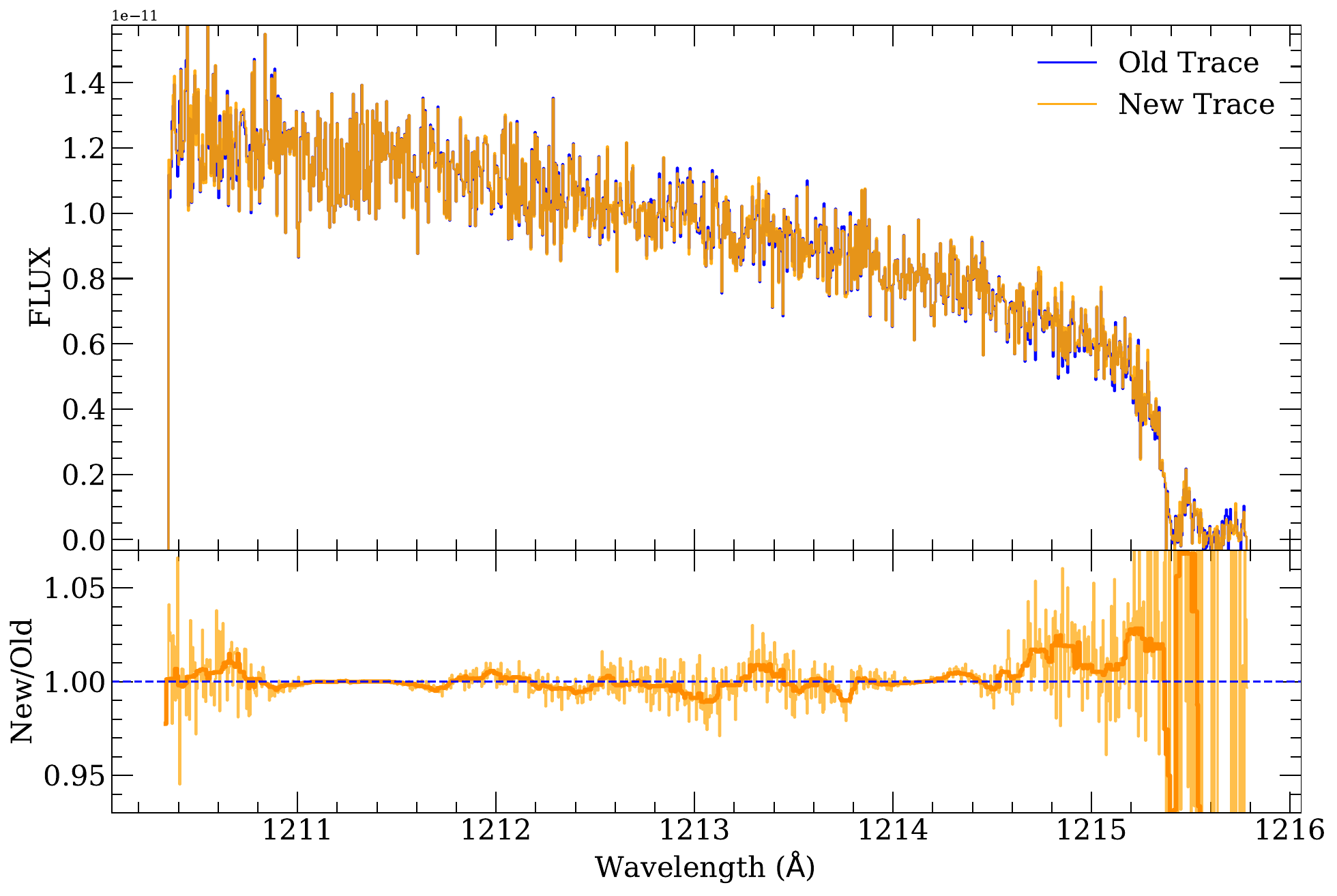}
    \caption{Example trace fit (top), improvements (E140H/1271 order 347) in extraction region (middle), gross counts (bottom-left) and flux (bottom-right) when reducing data using the previous straight-line secondary traces (blue curves) and our new Gaussian process-defined traces (orange curves). In this case, strong Lyman-$\alpha$ absorption lands on the edge of the detector. The trace is automatically extrapolated in this region via the Gaussian-process fit.}
    \label{fig:improve2}
\end{figure}

\end{document}